\documentclass[11pt]{article}
\pdfoutput=1
\usepackage[margin=1.0in]{geometry}
\usepackage{indentfirst}
\usepackage{epsfig}
\usepackage{multicol}
\title{\textbf{Correlation Between GC-content and Palindromes in Randomly Generated Sequences and Viral Genomes}}
\author{Andrew Ninh\\
	\small \emph{Corresponding email: andrewninh@ieee.org}
}
\date{}
\begin{document}
\maketitle

\begin{abstract}
GC-content, the ratio of guanine and cytosine bases in an entire nucleotide sequence, and palindromic sequences are unique for every organism due to genomic evolution. The goals of our research was to establish a correlation between GC-content and palindromic densities in wild-type viral and randomly-generated genomes. Forty viral genomes were downloaded from GenBank and their GC-ratios and palindromic densities were calculated and plotted using Mathematica. The palindromic densities-by-GC-ratios plot of randomly generated sequences (palindromic density curve) exhibited a quadratic relationship and was superimposed over the viral genome plot. It was observed that the viral plots followed the curvature of the random sequences' quadratic curve, signifying a directly proportional relationship between GC-content and palindrome density in viral genomes. However, because viral genomes require certain non-palindromic sequences to function, the palindromic densities of most wild-type genomes were under the palindromic density curve. The variance in palindrome densities of wild-type genomes in respect to the random sequences' quadratic curve may be examined to determine evolutionary traits in genomes. A better understanding of viral palindromic densities and GC-ratios would help in understanding conserved secondary RNA structures in viral genomes and future drug discovery. In addition, certain viral genomes were found to be viable recombinant viruses, which are used in gene therapy.
\end{abstract}

\begin{center}
\line(1,0){250}
\end{center}

\begin{multicols}{2}
\section{Introduction}
GC-content (guanine-cytosine content) refers to the percentage of a nucleotide sequence that is made up of either guanine or cytosine bases. Sequences with a high GC-content tend to be more stable than sequences with lower GC-content due to stacking interactions | not the fact that the GC pair has three hydrogen bonds and the AT pair has two (Yakovchuk et al., 2006). Interestingly, higher GC-content is speculated to be associated with autolysis (or cell destruction by its own enzymes) which reduces cell longevity despite genetic thermostability (Levin \& Van Sickle, 1976).

	GC-content varies throughout a genome and this variation is speculated to be driven by both selective and neutral processes (Pozzoli et al., 2008; Nishida, 2012). Genetic recombination and genomic evolution is also directly affected by silent GC-content (Birdsell, 2002). Hence, low GC-content and AT mutational bias is characteristic of nonrecombining genomes. Due to the codon usage bias in shorter sequences (due to the fact that the stop codon is a higher AT bias), shorter sequences tend to have lower GC-contents while longer sequences have higher GC-contents (Wuitschick \& Karrer, 1999).
	
	Due to the variation of GC-content in organisms, GC-content has been used as a method of classifying bacteria in higher level hierarchial classification (Wayne et al., 1987).

	DNA palindromes are inverted repeats that read identically from the 5' to 3' end as from the 3' to 5' end. These reverse complementary sequences, which examplify dyad symmetry, differ from lexical palindromes, which are the exactly identical forwards and backwards. The frequency of short palindromes has been found to be useful in comparative genomics by distinguishing and typing species (Lamprea-Burgunder et al., 2011).
	
	Protein folding rates have been found to be affected by both the GC-content of palindromes as well as palindromic density (Li \& Li, 2010). However, it is not known whether GC-content itself is correlated with the density or occurence of DNA palindromes.

\section{Materials and Methods}
Forty viral genomes of different lengths were chosen from information provided by the RNA virus database (Belshaw et al., 2008), the International Committee on Taxonomy of Viruses database (ICTVdB) (Fauquet \& Fargette, 2005), and the ViralZone database (Hulo et al., 2011). All viral genomes were downloaded from NCBI GenBank in FASTA file format (Benson et al., 2009). Their GC-contents and palindromic densities were calculated using a Java program (see Table 2).

	A program was written using the Java programming language to randomly generate genetic sequences based on a GC-ratio input. Because the \verb|Random| package provided by Oracle is not cryptographically secure, the \verb|SecureRandom| package was used to randomly generate sequences. The GC-content and palindromic density of randomly generated sequences were used as controls to compare randomly-generated sequences with wild-type viral genomes. 
	
			The GC-content of a DNA sequence is calculated by the sum of G and C bases divided by the total number of nucleotides.
$$GC\% = \frac{G + C}{A + T + G + C} \times 100$$
The resultant number is known as the GC-ratio of the sequence. Another program was written to calculate the GC-ratios of all randomly generated sequences and viral genomes.
	
	A Java program was also written to count the occurence of perfect (unbroken) DNA palindromes and divide that number by the length of the input sequence (giving a palindrome density) in viral genomes and randomly generated sequences. The palindrome density value is defined by the equation:
$$D(S) =  \frac{p(S) = \{ \sum P(S),\ \ell(P(S)) \geq 4 \}}{\ell (S)}$$
for which $S$ is any nucleotide sequence, $p(S)$ represents the sum of palindromes found by the function $P(S)$ (which is the Java program written to search for all palindromic sequences), $\ell (P(S))$ is the function that returns the length of a palindrome found by function $P(S)$ (to determine whether the palindrome is to be summed), and $\ell (S)$ represents the length as a function of the inputted sequence. Palindromes with a length of 4 are considered to be the shortest palindromes because sequences lesser than 4 would either be nucleotides or insignificant.

	It is important to note that this equation to find ``palindromic density'' provides an arbitrary constant that simply describes the ratio of certain-length palindromes to the length of a sequence. The $P(S)$ algorithm is a computational program written in the Java programming language to find all palindromes in a given sequences. This includes finding palindromes within palindromes. Although this would seem to overestimate the number of palindromic sequences, it gives a better account for palindrome lengths | a factor not included in other palindrome-detecting algorithms which cannot differentiate a long palindrome from a shorter palindrome.
		
	The palindromic densities per GC-ratio of randomly generated sequences were calculated and plotted; there is a quadratic relationship between GC-content and palindromic densities | with the lowest palindromic density at 50\% and the highest palindromic densities at the extremas. Random sequences, by their nature, exhibit perfect palindromic density relationships with GC-content whereas wild-type sequences have specific GC-contents and palindromic sequences (some of which are necessary for the genome to function). 
		
	In case of the probability of random chance occurences caused by the \verb|SecureRandom| package, 100 sequences were randomly generated per inputted length and their GC-content and palindromic densities were calculated. By the law of large numbers, the average of each trial's GC-content and palindromic density would be close to a stable number.

\section{Results}
Randomly generated sequences with a GC-content of 50\% demonstrated the lowest number of palindromic sequences whereas sequences with GC-contents of 25\% and 75\% had equally higher numbers of palindromes (see Table 1).

	The Mathematica plots of the random sequences' palindromes demonstrated that the palindromic sequences increased linearly as the length of randomly generated sequences increased (see Figure 1). The 25\% and 75\% plots exhibit identical slopes while the 50\% plot has a more gradual plot with a smaller slope.
	
	GC-contents and palindromic densities of randomly generated sequences exhibit a quadratic relationship (see Figure 2). Using the Mathematica \verb|Fit| function, this quadratic relationship is defined by the equation:
$$f(x) = 1.403 - 4.137 x + 4.132 x^2$$
Although the quadratic representation has maximas at 0 and 1 (representing GC-ratios of 0\% and 100\% respectively), wild-type genomes rarely, if ever, reach GC-ratios of lesser than 25\% or greater than 75\%. The minima of the equation, 0.5 (GC-content of 50\%), exhibits the lowest palindromic density.
	
	Forty viral genomes were downloaded from GenBank in FASTA file format through the previously stated databases and their GC-contents and palindromic densities were calculated. These values became ordered pairs (GC-content, palindromic density) and saved on a comma-separate values (csv) file which was plotted on Mathematica. Palindromic densities (ordinate value) by GC-contents (abscissa value) were plotted and superimposed over the palindromic density by GC-content graph of randomly generated sequences (see Figure 2). 
	
	It was observed that the majority of the GC-ratios of viral genomes were in the 40\% range and almost all viral palindromic densities were below the curve of the random GC-content-palindromc ratio plot.

\section{Discussion}
The lowest number of palindromc sequences can be found in the randomly generated sequences that have a GC-content of 50\%. The number of palindromes increases as GC-content rises and falls. This is most likely due to the fact that less bases, thus less available combinations of bases, will allow for more palindromes.

	As demonstrated in by the viral genomes, there is a slight correlation between GC-content and the presence of palindromic sequences. However, this relationship between GC-content and palindromic density varies per viral genome and there will always be outliers. This is because certain palindromes (such as restriction enzyme or methylation sites) are necessary for an organism to function. Although GC-content and palindromes are used separately to differentiate species, GC-content seems to dictate the palindromic density of most viral genomes, evidenced by how viral points on Figure 2 seem to follow the random palindrome density curve albeit below it.
	
	Palindromic density is also directly affected by the presence of non-palindromic sequences such as microsatellites or sequence motifs that, as aforementioned, code for proteins that are necessary in the functioning of wild-type genomes. These DNA motifs and naturally-occurring sequences are not accounted for in randomly generated sequenecs. This accounts for why almost all viral palindrome densities exist under the random palindrome density curve. 
	
	The degree at which certain viral genomes diverge from the random sequences' quadratic curve may provide details to a genomes' evolution, usage of palindromes (such as for methylation sites), and secondary structure stability. There has been record of cytosine methylation in mammalian DNA viruses, so viral genomes with a high GC-content and high palindrome density would have a high presence of CpG islands (sites of methylation) (Hoelzer et al., 2008).
	
	However, it seems that there are certain sequences whose GC-content has no affect on palindrome density. This occurence is most likely due to an increased number of necessary non-palindromic sequences. These genomes may either still exhibit conserved palindromes which lead to conserved secondary RNA structures, open reading frames, (non)coding regions in introns or exons of sequences (which may be highly conserved structures as well), or viral microsatellites. Understanding microsatellites and their polymorphisms may be important in differentiating different strains of viruses | particularly herpes simplex strains (Deback et al., 2009).
	
	As observed, the majority of the sequences contain a GC-ratio around the 40\% to 50\% range. In such a situation, these genomes are described as being slightly AT-rich rather than GC-poor (Musto et al., 1997). The lower GC-content is due to the fact that some genomes are short (characteristic of many viruses) and the stop codon of short genomes has an AT-bias. In addition, low GC-content is characteristic of nonrecombining genomes, meaning that viral genomes with higher GC-content may be used as recombinant viruses for viral vaccines or gene therapy.
	
	Palindromes are also necessary in the formation of secondary structures | many of which are conserved throughout viral genomes (Fekete, 2000; Hofacker et al., 2004). Understanding of conserved secondary structures in viral genomes would help in establishing the evolution of these small genomes and would possibly lead to vaccine discovery to target these conserved sequences.
	
	For future research, the palindrome-finding algorithm can be expanded to search for palindromic sequenecs with spacer sequences in addition to perfect palindromes. In addition, further research may be done in comaring the randomly generated GC-contents and palindrome densities (which are unchanging) to the genomes or genes of other species or more complex organisms.
	
\section{Conclusion}
GC-content and palindromic density share a non-linear, quadratic relationship with the lowest palindromic density at a GC-ratio of 50\% and higher palindromic densities as GC-ratios approach extreme values. 

	Most viral palindromic densities followed beneath the random palindrome density curve. This is due to the fact that viral genomes require non-palindromic sequences such as open reading frames (ORF) or splicing sites which are necessary for the natural functioning of a genome and would not exist in randomly generated sequences.
	
	In addition, the palindromic densities of viral genomes may be used to trace the evolution of viruses and through the understanding of conserved structures such as palindromes, future drug discoveries may use this to target certain viral genomes.
	
	Viral genomes with a high palindrome density relative to the random palindrome density curve and a higher GC-ratio ($>$50\%) may be used as recombinant viruses for future vaccines and gene therapy.

\end{multicols}

\begin{center}
\line(1,0){250}
\end{center}

\section{Tables and Figures}

\begin{table}[ht]
	\caption{Randomly Generated Sequences with Corresponding Palindrome Count by GC-Content}
	\centering
	\begin{tabular}{c c c c c c c}
		\hline \hline
		Length & & GC-content 25\% & & GC-content 50\% & & GC-content 75\% \\ [1ex]
		\hline
		500 & & 293 & & 208 & & 298 \\ [1ex]
		1000 & & 590 & & 418 & & 595 \\ [1ex]
		1500 & & 890 & & 624 & & 896 \\ [1ex]
		2000 & & 1189 & & 829 & & 1194 \\ [1ex]
		3000 & & 1802 & & 1252 & & 1782 \\ [1ex]
		5000 & & 2974 & & 2080 & & 2986 \\ [1ex]
		7500 & & 4466 & & 3124 & & 4482 \\ [1ex]
		10000 & & 5989 & & 4181 & & 5950 \\ [1ex]
		15000 & & 8953 & & 6271 & & 8973 \\ [1ex]
		20000 & & 11931 & & 8326 & & 11914 \\ [1ex]
		\hline
	\end{tabular}
	\label{table:hPal}
\end{table}

\begin{table}[ht]
	\caption{Viral Genomes with Corresponding Palindrome Count and GC-Content}
	\centering
	\scalebox{0.8}{
	\begin{tabular}{c c c c c c c c c}
		\hline \hline
		Species & & RefSeq & & Length & & GC-content & & Palindrome Density \\ [1ex]
		\hline
		Eyach Virus VP12 Protien & & NC\_003707.1 & & 678 & & 47.198\% & & 0.411 \\ [1ex]
		Rotavirus A VP7 & & NC\_011503.2 & & 1062 & & 35.782\% & & 0.495 \\ [1ex]
		Hepatitis Delta Virus & & NC\_001653.2 & & 1682 & & 58.799\% & & 0.335 \\ [1ex]
		Hantaan Virus & & NC\_005218.2 & & 1696 & & 42.935\% & & 0.384 \\ [1ex]
		H1N1 Hemagglutinin & & NC\_002017.1 & & 1778 & & 41.620\% & & 0.402 \\ [1ex]
		Rotavirus B VP2 & & NC\_007549.1 & & 2969 & & 34.321\% & & 0.480 \\ [1ex]
		Hepatitis B & & NC\_003977.1 & & 3215 & & 49.269\% & & 0.365 \\ [1ex]
		Adeno-associated Virus - 2 & & NC\_001401.2 & & 4679 & & 53.794\% & & 0.428 \\ [1ex]
		Enterobacteria Phage $\Phi$X174 & & NC\_001422.1 & & 5386 & & 44.764\% & & 0.396 \\ [1ex]
		Maize Rayado Fino Virus & & NC\_002786.1 & & 6305 & & 61.982\% & & 0.391 \\ [1ex]
		Turnip Yellow Mosaic Virus & & NC\_004063.1 & & 6318 & & 56.442\% & & 0.352 \\ [1ex]
		Human Astrovirus & & NC\_001943.1 & & 6813 & & 44.841\% & & 0.386 \\ [1ex]
		Human Rhinovirus A 89 & & NC\_001617.1 & & 7152 & & 39.038\% & & 0.412 \\ [1ex]
		Hepatitis E & & L08816.1 & & 7176 & & 57.943\% & & 0.458 \\ [1ex]
		Grapevine Fleck Virus & & N\_003347.1 & & 7564 & & 66.235\% & & 0.316 \\ [1ex]
		Foot-and-mouth Diseas Virus Type O & & N\_004004.1 & & 8134 & & 55.274\% & & 0.392 \\ [1ex]
		Cassava Vein Mosaic Virus & & N\_001648.1 & & 8159 & & 24.930\% & & 0.504 \\ [1ex]
		Aichi Virus & & NC\_001918.1 & & 8251 & & 58.902\% & & 0.358 \\ [1ex]
		Human T-lymphotropic Virus 1 & & NC\_001436.1 & & 8507 & & 53.462\% & & 0.404 \\ [1ex]
		Human Immunodificiency Virus 1 & & NC\_001802.1 & & 9181 & & 42.120\% & & 0.401 \\ [1ex]
		GB Virus C & & NC\_001710.1 & & 9393 & & 59.082\% & & 0.420 \\ [1ex]
		Hepatitis C & & NC\_004102.1 & & 9464 & & 58.231\% & & 0.414 \\ [1ex]
		Rubella Virus & & NC\_001545.2 & & 9762 & & 69.596\% & & 0.505 \\ [1ex]
		Human Immunodificiency Virus 2 & & NC\_001722.1 & & 10351 & & 45.661\% & & 0.382 \\ [1ex]
		Louping Ill Virus & & NC\_003690.1 & & 10871 & & 54.852\% & & 0.378 \\ [1ex]
		Langat Virus & & NC\_003690.1 & & 10943 & & 54.309\% & & 0.377 \\ [1ex]
		Rabies Virus & & NC\_001542.1 & & 11932 & & 45.097\% & & 0.386 \\ [1ex]
		Japanese Encephalitis Virus & & NC\_001437.1 & & 10976 & & 51.421\% & & 0.364 \\ [1ex]
		Mayaro Virus & & NC\_003417.1 & & 11411 & & 50.372\% & & 0.411 \\ [1ex]
		Simian Foamy Virus & & NC\_001364.1 & & 13246 & & 38.955\% & & 0.430 \\ [1ex]
		Human Respiratory Syncytial Virus & & NC\_001781.1 & & 15225 & & 47.427\% & & 0.402 \\ [1ex]
		Human Parainfluenza Virus 1 & & NC\_003461.1 & & 15600 & & 33.557\% & & 0.449 \\ [1ex]
		Measles Virus & & NC\_001498.1 & & 15895 & & 47.427\% & & 0.402 \\ [1ex]
		Nipah Virus & & NC\_002728.1 & & 18246 & & 38.167\% & & 0.446 \\ [1ex]
		Ebola Virus & & NC\_002549.3 & & 18959 & & 41.073\% & & 0.441 \\ [1ex]
		Marburg Virus & & NC\_001608.3 & & 19111 & & 38.292\% & & 0.440 \\ [1ex]
		Acidianus Bottle-shaped Virus & & NC\_009452.1 & & 23814 & & 34.564\% & & 0.429 \\ [1ex]
		Gill-associated Virus & & NC\_010306.1 & & 26253 & & 46.235\% & & 0.385 \\ [1ex]
		Human Coronavirus Virus NL63 & & NC\_005831.2 & & 27553 & & 34.461\% & & 0.429 \\ [1ex]
		SARS Coronavirus & & NC\_004718.3 & & 29751 & & 40.672\% & & 0.414 \\ [1ex]
		\hline
	\end{tabular}}
	\label{table:hPal}
\end{table}

\begin{figure}[ht]
\centering
\includegraphics[height=100mm]{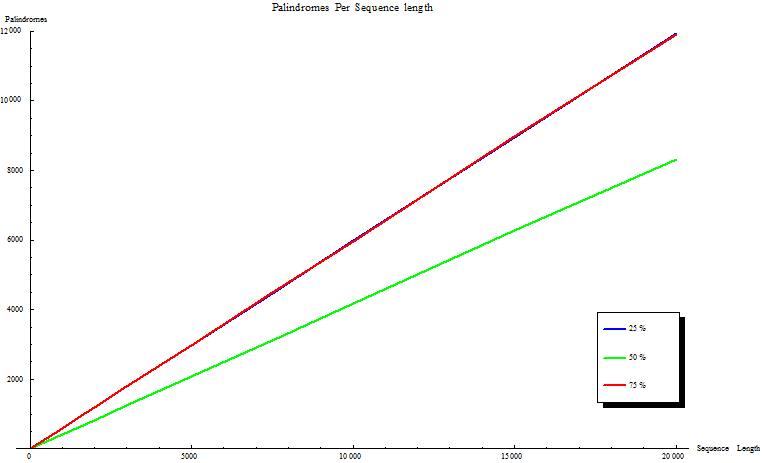}
\caption{
	\small
	A plot of the number of palindromes by length of randomly generated sequences with GC-contents of 25\%, 50\%, and 75\%.
}
\end{figure}

\begin{figure}[ht]
\centering
\includegraphics[height=100mm]{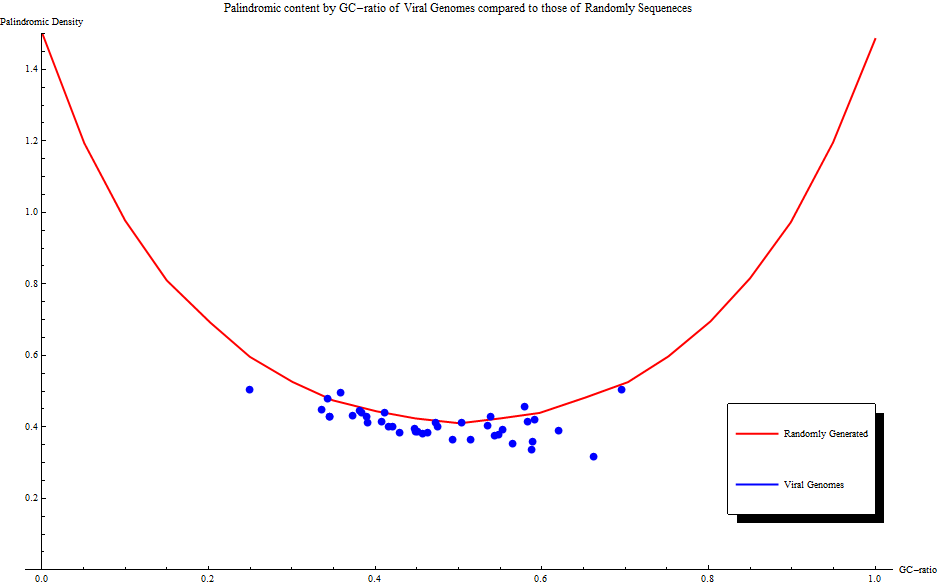}
\caption{
	\small
	A plot demonstrating the relation of Palindromc Density and GC-ratio of viral genomes compared to that of randomly generated sequences.
}
\end{figure}

\end{document}